\documentstyle[prb,aps,epsfig,multicol]{revtex}
\begin{document}
\def\etal{{\it et al.}}
\def\LJ#1{$\rm LJ_{#1}$}
\def\csix#1{$ ({\rm C}_{60})_{#1}$}
\def\csixty{$ {\rm C}_{{60}}$}

\title{Global Optimization by Basin-Hopping and the Lowest Energy
Structures of Lennard-Jones Clusters Containing up to 110 Atoms}
\author{David J.~Wales}
\address{University Chemical Laboratories, Lensfield Road, Cambridge CB2 1EW, UK}
\author{Jonathan P.~K.~Doye}
\address{FOM Institute for Atomic and Molecular Physics, Kruislaan 407,\\
1098 SJ Amsterdam, The Netherlands}
\maketitle
\begin{abstract}
We describe a global optimization technique using \lq basin-hopping' in 
which the potential energy surface is transformed into a collection of interpenetrating
staircases. This method has been designed to exploit the features which recent work
suggests must be present in an energy landscape for efficient relaxation to
the global minimum.
The transformation associates any point in configuration space with the local 
minimum obtained by a geometry optimization started from that point, effectively
removing transition state regions from the problem. However, unlike other methods
based upon hypersurface deformation, this transformation does not change the
global minimum. The lowest known structures are located for all Lennard-Jones clusters 
up to 110 atoms, including a number that have never been found before in unbiased searches.
\end{abstract}  
\begin{multicols}{2}
\section{Introduction}Global optimization is a subject of intense current interest.\cite{SA}
Improved global optimization methods could be of great economic importance, since improved solutions
to travelling salesman type problems, the routing of circuitry in a chip, the active structure of a 
biomolecule, etc., equate to reduced costs or improved performance. 
In chemical physics the interest in efficient global optimization methods stems from the common problem
of finding the lowest energy configuration of a (macro)molecular system.
For example, it seems likely that the native structure of a protein is structurally related to the
global minimum of its potential energy surface (PES). If this global minimum could be found reliably
from the primary amino acid sequence, this knowledge
would provide new insight into the nature of protein folding and save biochemists many
hours in the laboratory. Unfortunately, this goal is far from being realized. 
Instead the development of global optimization methods has usually concentrated on much simpler systems.

Lennard-Jones (LJ) clusters represent one such test system.
Here the potential is
\begin{equation}
E = 4\epsilon \sum_{i<j}\left[ \left(\sigma\over r_{ij}\right)^{12} - \left(\sigma\over r_{ij}\right)^{6}\right], 
\end{equation}
where $\epsilon$ and $2^{1/6}\sigma$ are the pair equilibrium well depth and separation, respectively.
We will employ reduced units, i.e.~$\epsilon=\sigma=1$ throughout.
Much of the initial interest in LJ clusters was motivated by a desire to calculate nucleation rates
for noble gases.
However, as a result of the wealth of data generated, the LJ potential has been used not 
only for studying global optimization but also the effects of finite size
on phase transitions such as melting.
Through the combined efforts of many workers likely candidates
for the global minima of LJ$_N$ clusters have been found up to $N=147$.
\cite{HoareP71,HoareP71b,HoareP72,HoareP75,Hoare79,Freeman85,Farges85,Wille87,Northby87,Coleman,Xue,Pillardy,JD95c,JD95d,Deaven96}
This represents a significant achievement since
extrapolation of Tsai and Jordan's comprehensive enumeration of minima for small LJ clusters\cite{Tsai93a}
suggests that the PES of the 147-atom cluster possesses of the order of $10^{60}$ minima.\cite{pinch}

Previous studies have revealed that the Mackay icosahedron\cite{Mackay} provides the 
dominant structural motif for LJ clusters in the size range 10--150 atoms.
Complete icosahedra are possible at $N=13,55,147,\dots$ 
At most intermediate sizes the global minimum consists of a Mackay icosahedron at the core 
covered by a low energy overlayer. 
As a consequence of the phase behaviour of LJ clusters, 
finding these global minima is relatively easy.
Studies have shown that in the region of the solid-liquid transition the cluster is 
observed to change back and forth between a liquid-like form and icosahedral structures.\cite{Kunz93,Kunz94}
As a result of this `dynamic coexistence' a method as crude as molecular 
dynamics within the melting region coupled with systematic minimization of configurations
generated by the trajectory is often sufficient to locate the global minimum.\cite{dcoex}

\begin{figure}
\begin{center}
\vglue -0.2cm
\epsfig{figure=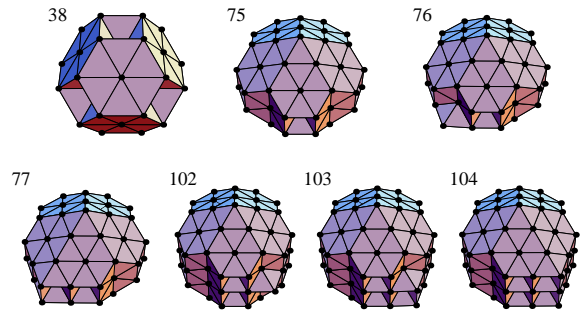,width=9.0cm}
\vglue -0.2cm
\begin{minipage}{8.5cm}
\caption{\label{fig:nonicos}Non-icosahedral Lennard-Jones global minima.}
\end{minipage}
\end{center}
\end{figure}

However, there are a number of sizes at which the global minimum is not based on an icosahedral structure.
These clusters are illustrated in Figure \ref{fig:nonicos}.
For \LJ{38} the lowest energy structure is a face-centred-cubic (fcc) truncated octahedron,\cite{Pillardy,JD95c}
and for $N=75$, 76, 77, 102, 103 and 104 geometries based on Marks' decahedra\cite{Marks84}
are lowest in energy.\cite{JD95c,JD95d}
For these cases finding the lowest minimum is much harder because the global minimum of free energy only
becomes associated with the global potential energy minimum at temperatures well below melting where the dynamics of 
structural relaxation are very slow.
For \LJ{38}, the microcanonical temperature for the transition from face-centred cubic
to icosahedral structures has been estimated to be about 0.12$\epsilon k^{-1}$,
where $k$ is the Boltzmann constant,
and for \LJ{75} the estimate for the decahedral to icosahedral transition 
is about 0.09$\epsilon k^{-1}$.\cite{hsm}
(For comparison, melting typically occurs at about $T=0.2-0.3\epsilon k^{-1}$.)

The topography of the PES can also play a key role in determining the ease of global optimization\cite{JD96c}.
A detailed study of the \LJ{38} PES has shown that there is a large energy barrier
between the fcc and icosahedral structures\cite{JD97a} which correspond to well-separated 
regions of the PES.
Furthermore, fcc and decahedral structures have less polytetrahedral character than 
icosahedral structures, and hence they have less in common with the liquid-like state, which is
characterized by disordered polytetrahedral packing.\cite{NelsonS,JD96a,JD96b}
Since the vast majority of configuration space is dominated by `liquid-like' configurations,
it is therefore harder to find global minima based upon fcc and decahedral packing using
unbiased searches.

These considerations explain why global optimization methods have only recently begun to find
the truncated octahedron\cite{Pillardy,Deaven96,Niesse96a,Barron96}
and why, until now, the Marks' decahedron has never been found by an 
unbiased global optimization method.
The greater difficulty of finding the \LJ{75} global minimum compared to \LJ{38} can 
probably be explained by the slightly smaller transition temperature, 
the sharper transition and the much larger number of minima on the \LJ{75} PES.

Before we consider the effectiveness of different global optimization methods for
Lennard-Jones clusters it is interesting to note that the use of physical principles 
to construct good candidate structures for the global 
minima\cite{HoareP71,HoareP71b,HoareP72,HoareP75,Farges85,JD95c,JD95d} 
or to reduce the configuration space that needs to be searched\cite{Northby87,Coleman,Xue} led to the 
initial discovery of 93\% of the LJ global minima in this size range.
It seems that physical insight into a specific problem 
will often be able to beat unbiased global optimization, a view expressed by Ngo \etal\cite{Ngo94}
in their discussion of computational complexity.

One difficulty in evaluating the relative performances of different global optimization methods is
that, too often, the methods have only been applied to small clusters, or to larger clusters
with global minima that are especially stable, such as \LJ{55}.
It is also difficult to draw any firm conclusions about how efficient different methods
may be when the number of searches employed varies widely. However,
it seems reasonable to suggest two hurdles that any putative global optimization approach
should aspire to. The first is the location of the truncated octahedron for \LJ{38}, and any
method which fails this test is unlikely to be useful. The second hurdle is the location of the
Marks decahedron for \LJ{75}; this problem poses a much more severe test for an unbiased 
search and one which does not appear to have been passed until the present work.

The most successful global optimization results for LJ clusters reported to date are
for genetic 
algorithms.\cite{Deaven96,Niesse96a,Gregurick,Mestres}
These methods mimic some aspects of biological evolution: a population of clusters evolves to 
low energy by mutation and mating of structures, and selection of those with low potential energy.
To be successful new configurations produced by `genetic manipulation' are mapped 
onto minima by a local optimization algorithm such as the conjugate gradient method.
The study by Deaven \etal\ is particularly impressive, since these workers matched or beat all the lowest
energy minima that they knew of up to $N=100$, including the truncated octahedron 
(although they probably missed the global minima at $N$=69 and 75--78).
Niesse and Mayne were also able to locate the \LJ{38} truncated octahedron, and report that
this structure took about 25 times longer to find than the icosahedral global minima 
of the neighbouring sizes.

Another class of global optimization techniques, sometimes called hypersurface deformation methods, 
attempts to simplify the problem by applying a transformation to the PES which smoothes it 
and reduces the number of local minima.\cite{StillW88,Piela94}
The global minimum of the deformed PES is then mapped back to the original surface
in the {\it hope} that this will lead back to the global minimum of the original PES.
The distinctions between the various methods of this type lie in the type of transformations that are used,
which include applying the diffusion equation,\cite{Kostrowicki,Scheraga92}
increasing the range of the potential,\cite{StillD90b,Head91}
and shifting the position of the potential minimum towards the origin.\cite{Pillardy92}
The performance of hypersurface deformation methods has been variable: the potential shift approach
managed to find the 38-atom truncated octahedron, but other workers report difficulties\cite{Kostrowicki,Scheraga92} for
the trivial cases of \LJ{8} and \LJ{9} where
there are only 8 and 21 minima on the PES, respectively.

Although intuitively appealing, the problem with hypersurface deformation is that
there is no guarantee that the global minimum on the deformed PES will
map onto the global minimum of the original surface. 
This difficulty is clearly illustrated when we consider Stillinger and Stillinger's 
suggestion of increasing the range of the potential:\cite{StillD90b}
it has been shown that the global minimum may in fact depend rather sensitively
on the range of the potential, with the appearance of numerous `range-induced' 
transitions.\cite{JD95c,JD97e}

Other methods include those based on `annealing'.
Such approaches take advantage of the simplification in the free energy landscape 
that occurs at high temperatures,
and attempt to follow the free energy global minimum as the temperature is decreased.
At zero Kelvin the free energy global minimum and the global minimum of the PES must coincide. 
Standard simulated annealing\cite{KirkSA} was used by Wille to find a few new minima at small sizes\cite{Wille87}
but does not appear to have been systematically applied to LJ clusters.
More sophisticated variants of this technique include
gaussian density annealing and analogues,\cite{Ma93,Ma,Tsoo94,Schelstrate97}
but again some appear to fail at small sizes.\cite{Tsoo94,Schelstrate97}

The difficulty with the annealing approach methods is that, if the free energy global minimum changes at low 
temperatures where dynamical relaxation is slow, 
the algorithms will become stuck in the structure corresponding to the high temperature free energy global minimum.
Such methods are therefore likely to experience difficulties in finding
the global minima for \LJ{38} and \LJ{75}. In the language employed in recent protein folding
literature,\cite{Bryngel95} annealing will fail when $T_f < T_g$, where $T_f$ is the `folding' temperature
below which the global potential energy and free energy minima coincide, and $T_g$ is the
`glass' temperature at which the system effectively becomes trapped in a local minimum.

Another method which attempts to reduce the effects of barriers on the PES makes use of quantum tunnelling.
The diffusion Monte Carlo approach is used to find the ground state wavefunction, 
which should become localized at the global minimum as $\hbar$ is decreased to zero.\cite{Finnila96}
A more rigorous approach has been applied by Maranas and Floudas, who found upper and lower
bounds for the energy of the global minimum. However, the computational expense of this method,
which scales as $2^N$ with the number of atoms, means that it has only been used for small
systems.\cite{Maranas92,Maranas94}
Most of the above studies, along with the recently described `pivot method'\cite{Stanton97,Serra97a,Serra97b}
and `taboo search',\cite{Cvijovic,Hong97} have yet to prove their usefulness 
by passing the first hurdle for \LJ{38} suggested above.
However, this does not necessarily mean that these approaches should be discounted, since
some authors have only applied their algorithms to smaller clusters and may not have run enough searches
to achieve convergence. 

In the present work
we present the results of a \lq basin-hopping' global optimization technique for Lennard-Jones clusters. 
All the known lowest energy structures up to $N=110$ have been located successfully, including 
three minima not previously reported. (See Tables 1 and 2.)
This method is also the first unbiased algorithm to find the global minima based on the Marks decahedron 
around \LJ{75}\ and \LJ{102}.
For reference, we collect the rather scattered results previously reported for LJ clusters to 
provide a complete catalogue of the energies and point groups of the lowest energy minima that we know of.
The results have been collected in the first entry of the Cambridge Cluster Database
at http://brian.ch.cam.ac.uk.

\section{Method}
\begin{figure}
\begin{center}
\epsfig{figure=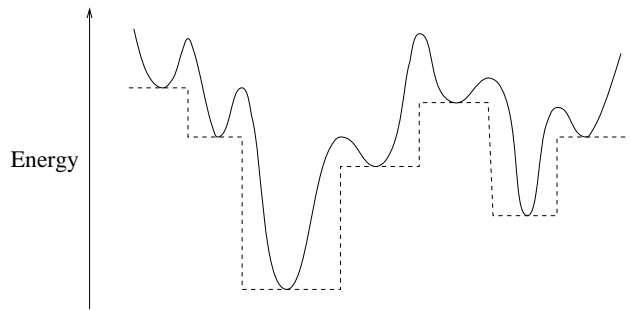,width=8.2cm}
\vglue 0.2cm
\begin{minipage}{8.5cm}
\caption{\label{fig:trans}
A schematic diagram illustrating the effects of our energy transformation 
for a one-dimensional example. 
The solid line is the energy of the original surface and the dashed line is the 
transformed energy, $\tilde E$.}
\end{minipage}
\end{center}
\end{figure}

The present approach has been guided by previous work on energy landscapes which has
identified features that enable the system to locate its global minimum efficiently.\cite{Bryngel95}
In particular, analysis of model energy landscapes using a master equation approach for
the dynamics, has provided good evidence that such a surface should have a large potential
energy gradient and the lowest possible transition state energies or rearrangement barriers.\cite{JD96c}
These results immediately suggest a simple way to transform the PES which does not change
the global minimum, nor the relative energies of any local minima. We consider the 
transformed energy $\tilde E$ defined by:
\begin{equation}
\tilde E({\bf X}) = min\left\{ E({\bf X}) \right\}, 
\end{equation}
where ${\bf X}$ represents the $3N$-dimensional vector of nuclear coordinates and
$min$ signifies that an energy minimization is performed starting from ${\bf X}$.
In the present work energy minimizations were performed using the Polak-Ribiere variant
of the conjugate gradient algorithm.\cite{Recipes}
Hence the energy at any point in configuration space is assigned to that of the local
minimum obtained by the given geometry optimization technique, and the PES is mapped
onto a set of interpenetrating staircases with plateaus corresponding to the set of
configurations which lead to a given minimum after optimization. 
A schematic view of the staircase topography that results from this transformation 
is given in Figure \ref{fig:trans}. 
These plateaus, or basins of attraction, have been visualized in previous work as a means to compare the
efficiency of different transition state searching techniques.\cite{Wales92,Wales93c}

\vglue -0.2cm

The energy landscape for the function $\tilde E({\bf X})$ was explored using a 
canonical Monte Carlo simulation at a {\it constant\/} reduced temperature of $0.8$. 
At each step all coordinates were displaced by a random number in the range
$[-1,1]$ times the step size, which was adjusted to give an acceptance ratio
of $0.5$. The nature of the transformed surface allowed relatively large step
sizes of between $0.36$--$0.40$. For each cluster in the range considered seven
separate runs were conducted. Five of these each consisted of 5000 Monte Carlo
steps starting from different randomly generated configurations of atoms confined
to a sphere of radius $5.5$ reduced units. The subsequent geometry optimizations
employed a container of radius one plus the value required to contain the same
volume per atom as the fcc primitive cell. The container should have little effect
on any of our results and is only required to prevent dissociation during the
conjugate gradient optimizations.

The convergence criterion employed for the conjugate gradient optimizations used
in the Monte Carlo moves need not be very tight. In the present work we required
the root-mean-square (RMS) gradient to fall below $0.01$ in reduced units and the energy
to change by less than $0.1\,\epsilon$ between consecutive steps in the
conjugate gradient search. Initially it appeared that a tolerance of $0.1$ for the
RMS gradient was satisfactory, but this was subsequently found to cause problems
for clusters containing more than about 60
atoms. The lowest energy structures obtained during the canonical simulation were saved
and reoptimized with tolerances of $10^{-4}$ and $10^{-9}$ for the RMS force and the
energy difference, respectively. The final energies are accurate to about six decimal
places.

Several other techniques were employed in these calculations, namely seeding, freezing and angular moves.
Here we used the pair energy per atom, $E(i)$, defined as
\begin{equation}
E(i) = 4\epsilon \sum_{j\not=i}\left[\left(\sigma\over r_{ij}\right)^{12}-\left(\sigma\over r_{ij}\right)^6\right],
\end{equation}
so that the total energy is
\begin{equation}
E = {1\over2} \sum_i E(i).
\end{equation}
If the highest pair energy rose above a fraction $\alpha$ of the lowest pair energy
then an angular move was employed for the atom in question with all other atoms fixed.
$\alpha$ was adjusted to give an acceptance ratio for angular moves of $0.5$ and
generally converged to between $0.40$ and $0.44$. Each angular displacement consisted of choosing
random $\theta$ and $\phi$ spherical polar coordinates for the atom in question, taking the
origin at the centre of mass and replacing the radius with the maximum value
in the cluster.

The two remaining runs for each size consisted of only 200 Monte Carlo steps starting
from the global minima obtained for the clusters containing one more and one less
atom. When starting from \LJ{}$_{N-1}$ the $N-1$ atoms were frozen for the first 100 steps,
during which only angular moves were attempted for the remaining atom, starting from a
random position outside the core. When starting from \LJ{}$_{N+1}$ the atom with the highest
pair energy, $E(i)$, was removed and 200 unrestricted Monte Carlo moves were attempted
from the resulting geometry.

The above basin-hopping algorithm shares a common philosophy with our previous approach 
in which steps were taken directly between minima using eigenvector-following to
calculate pathways.\cite{JD97a}
The latter method is similar to that described recently by Barkema and Mousseau\cite{Barkema96a}
in their search for well-relaxed configurations in glasses. 
Although the computational expense of transition state searches probably makes this method 
uncompetitive for global optimization, our study illustrated the possible advantage
of working in a space in which only the minima are present.
The basin-hopping algorithm differs in that it is applied in configuration space
to a transformed surface, 
rather than in a discrete space of minima, and steps are taken stochastically.
The genetic algorithms described by Deaven \etal\cite{Deaven96} and Niesse and Mayne\cite{Niesse96a}
used conjugate gradient minimization to refine the local minima which 
comprise the population of structures that are evolved in their procedure. Hence these
authors are in effect studying the same transformed surface as described above, but explore
it in a rather different manner. We suspect that the success of their methods is at least
partly due to the implicit use of the transformed surface $\tilde E$.

The transformation of the PES also reduces the barriers to dissociation. Therefore, to prevent
evaporation either the cluster can be by placed in a tight-fitting container or 
the coordinates of the current point in configuration space can be reset to that 
of the minimum after each successful step. In this paper we used the latter method
and then the present approach is essentially the same as the 
\lq Monte Carlo-minimization' algorithm of Li and Scheraga,\cite{Li87a} 
who applied it to search the conformational space of the pentapeptide [Met$^5$]enkephalin. 
A similar method has recently been used by Baysal and Meirovitch\cite{Baysal96} to search the
conformational space of cyclic polypeptides.

\vglue -0.3cm
\begin{figure}
\begin{center}
\epsfig{figure=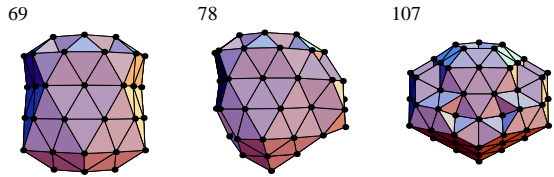,width=8.5cm}
\vglue -0.3cm
\begin{minipage}{8.5cm}
\caption{\label{fig:new}
Lennard-Jones global minima that have not previously been reported.}
\end{minipage}
\end{center}
\end{figure}

\section{Results}The basin-hopping algorithm has successfully located all the lowest known
minima up to \LJ{110}, including all the non-icosahedral structures illustrated in Figure \ref{fig:trans}
(sizes 38, 75, 76, 77, 102, 103 and 104)
and three new geometries based upon icosahedra illustrated in Figure \ref{fig:new} (sizes 69, 78 and 107). 
We believe that 
this is the first time any of the six decahedral global minima have been
located by an unbiased algorithm. The total number of searches was fixed in our
calculations to provide a simple reference criterion. In fact, most of the global
minima were found in more than one of the separate Monte Carlo runs. The global minima
for the smallest clusters were located within a few steps in each of the seven runs.
To give a better idea of how the algorithm performed we will provide some more details
for the sizes with non-icosahedral or newly discovered icosahedral global minima.

For \LJ{38} the truncated octahedron was found in four out of five of the longer unseeded
runs; the first success occurred within a thousand Monte Carlo steps on average. Not
surprisingly, the global minimum was not located in the shorter runs starting from the
structurally unrelated global minima for $N=37$ and 39.

For \LJ{75} the global minimum was found in just one of the longer Monte Carlo runs,
and again in the short run from the global minimum for \LJ{76}. However, the latter
minimum was only found in the short runs seeded from the \LJ{75} and \LJ{77} decahedra. 
Similarly, the \LJ{77} global minimum was only found in the short run seeded from the
\LJ{76} decahedron. 
The decahedral global minimum for \LJ{75} was found in four 
out of 100 Monte Carlo runs of 5000 steps each,
a frequency which fits in quite well with our
results for \LJ{75}, \LJ{76} and \LJ{77}. A successful run requires an initial geometry
which falls within the decahedral catchment area; all the other runs produce the lowest
icosahedral minimum after which the decahedron is never found. It would obviously be
possible to locate global minima based upon decahedra more efficiently by biasing the
starting configuration, but our intention was to analyze the performance of an unbiased
algorithm in the present work.

The pattern for \LJ{75}--\LJ{77} is repeated for \LJ{102}--\LJ{103}.
For \LJ{102} the decahedral global minimum was located in one of the longer Monte Carlo 
runs and in the short run seeded from the global minimum of \LJ{103}. The decahedral
minima for \LJ{103} and \LJ{104} were only found in short runs seeded from larger or smaller
decahedra. The decahedral global minimum for \LJ{102} was found in three
out of 100 Monte Carlo runs of 5000 steps each.

The three new icosahedral global minima all have an atom missing from a vertex
of the underlying Mackay icosahedron (Figure \ref{fig:new}). 
This is a possibility that Northby did not consider in his restricted
search of the icosahedral configuration space.
The new \LJ{69} global minimum was found in three of the five longer
Monte Carlo runs and in the short run seeded from \LJ{70}. The new global minimum
for \LJ{78} was only found in the short run seeded from \LJ{79}. The new minimum
for \LJ{107} was found in one of the longer Monte Carlo runs and in the short run
seeded from \LJ{108}.

We also performed a few preliminary runs for \LJ{192}\ and \LJ{201}, sizes at
which a complete Marks decahedron and a complete truncated octahedron occur,
respectively. For \LJ{192}\ the Marks decahedron has energy $-1175.697144$.
This structure was not found in 100 MC runs of 10000 steps each; instead the
lowest minimum located had an energy of $-1174.919801$. For \LJ{201}\ the
truncated octahedron has energy $-1232.731497$. However, we located a structure
of energy $-1236.124253$ which is based upon icosahedral packing. This minimum
was found in three out of 50 MC runs of 10000 steps each. For these larger systems greater 
efficiency could probably be achieved by varying the temperature and other 
parameters of the MC search.

\section{Conclusions}We have presented the results of a \lq basin-hopping' or
\lq Monte Carlo-minimization'\cite{Li87a} approach to global optimization
for atomic clusters bound by the Lennard-Jones potential containing
up to 110 atoms. All the lowest known minima were located successfully, including
the seven structures based upon fcc or decahedral packing and three new global
minima based upon icosahedra. Of the latter ten structures, only the smallest has
been located before by an unbiased algorithm, to the best of our knowledge.

The method is based upon a hypersurface deformation in which the potential energy
surface (PES) is converted into a set of plateaus each corresponding to a basin of 
attraction of a local minimum on the original PES.
This process removes all the transition state regions but does not affect the
energies of the minima. 
On the original PES, most trajectories that approach the boundary between
two basins of attraction are reflected back due to the high potential energy; 
only if the trajectory is along a transition state valley does passage between 
basins become likely. 
In contrast, on the transformed PES it is feasible for the system to hop 
between basins at any point along the basin boundary which dramatically 
reduces the time scale for interbasin motion.
We speculate that the success of a previous genetic algorithm
applied to the same clusters may be at least partly due to the fact that the same
space is implicitly considered in that approach.\cite{Deaven96}

The efficiency of the present approach could doubtless be improved by combining it with
various other techniques. The most obvious short-cut would be to start not from 
initial random configurations but from seeds with either decahedral, icosahedral 
or fcc morphologies. We have already checked that such biasing is indeed effective,
but our aim in the present paper was to gauge the performance of the unbiased algorithm.
The temperature at which our Monte Carlo runs were conducted was also not optimized
systematically.

Finally, as we noted in the introduction, global optimization for Lennard-Jones
clusters at most sizes is a relatively easy task. 
A more stringent and general test is provided
by Morse clusters which exhibit different structural behaviour as a 
function of the range of the potential.\cite{JD95c,JD97e} 
At short range the task is particularly difficult because the PES is very
rugged---the number of minima\cite{JD96b}
and the barrier heights\cite{Wales94b} increase as the range is decreased.

\end{multicols}
\begin{minipage}{17.cm}
\begin{table}
\caption{\label{table:gmin}Global minima of LJ$_N$ for $N\le 110$.
The references in which each minimum was first reported (to the best of our knowledge) are given,
and $\dagger$ indicates the present work. We intend to maintain an updated database of energies
and coordinates for LJ and Morse clusters on our web site: http://brian.ch.cam.ac.uk}
\begin{tabular}{ccccccccc}
 N & Point group & Energy/$\epsilon$ & Ref. & & N & Point group & Energy/$\epsilon$ & Ref. \\
\hline
 2 & $D_{\infty h}$ & -1.000000 & \onlinecite{HoareP71} & &	 57 & $C_s$ & -288.342625 & \onlinecite{Northby87} \\
 3 & $D_{3h}$ & -3.000000 & \onlinecite{HoareP71,HoareP71b} & &	 58 & $C_{3v}$ & -294.378148 & \onlinecite{Northby87} \\
 4 & $T_d$ & -6.000000 & \onlinecite{HoareP71,HoareP71b} & &	 59 & $C_{2v}$ & -299.738070 & \onlinecite{Northby87} \\
 5 & $D_{3h}$ & -9.103852 & \onlinecite{HoareP71,HoareP71b} & &	 60 & $C_s$ & -305.875476 & \onlinecite{Northby87} \\
 6 & $O_h$ & -12.712062 & \onlinecite{HoareP71,HoareP71b} & &	 61 & $C_{2v}$ & -312.008896 & \onlinecite{Northby87} \\
 7 & $D_{5h}$ & -16.505384 & \onlinecite{HoareP71,HoareP71b} & &	 62 & $C_s$ & -317.353901 & \onlinecite{Northby87} \\
 8 & $C_s$ & -19.821489 & \onlinecite{HoareP71,HoareP71b} & &	 63 & $C_1$ & -323.489734 & \onlinecite{Northby87} \\
 9 & $C_{2v}$ & -24.113360 & \onlinecite{HoareP71,HoareP71b} & &	 64 & $C_s$ & -329.620147 & \onlinecite{Northby87} \\
 10 & $C_{3v}$ & -28.422532 & \onlinecite{HoareP71} & &	 65 & $C_2$ & -334.971532 & \onlinecite{Xue} \\
 11 & $C_{2v}$ & -32.765970 & \onlinecite{HoareP71} & &	 66 & $C_1$ & -341.110599 & \onlinecite{Coleman,Xue} \\
 12 & $C_{5v}$ & -37.967600 & \onlinecite{HoareP71} & &	 67 & $C_s$ & -347.252007 & \onlinecite{Northby87} \\
 13 & $I_h$ & -44.326801 & \onlinecite{HoareP71,HoareP71b} & &	 68 & $C_1$ & -353.394542 & \onlinecite{Northby87} \\
 14 & $C_{3v}$ & -47.845157 & \onlinecite{HoareP71,HoareP71b} & &	 69 & $C_{5v}$ & -359.882566 & $\dagger$ \\
 15 & $C_{2v}$ & -52.322627 & \onlinecite{HoareP71} & &	 70 & $C_{5v}$ & -366.892251 & \onlinecite{Northby87} \\
 16 & $C_s$ & -56.815742 & \onlinecite{HoareP71} & &	 71 & $C_{5v}$ & -373.349661 & \onlinecite{Northby87} \\
 17 & $C_2$ & -61.317995 & \onlinecite{Freeman85} & &	 72 & $C_s$ & -378.637253 & \onlinecite{Coleman} \\
 18 & $C_{5v}$ & -66.530949 & \onlinecite{HoareP71} & &	 73 & $C_s$ & -384.789377 & \onlinecite{Northby87} \\
 19 & $D_{5h}$ & -72.659782 & \onlinecite{HoareP71} & &	 74 & $C_s$ & -390.908500 & \onlinecite{Northby87} \\
 20 & $C_{2v}$ & -77.177043 & \onlinecite{HoareP71} & &	 75 & $D_{5h}$ & -397.492331 & \onlinecite{JD95c} \\
 21 & $C_{2v}$ & -81.684571 & \onlinecite{HoareP71} & &	 76 & $C_s$ & -402.894866 & \onlinecite{JD95c} \\
 22 & $C_s$ & -86.809782 & \onlinecite{Northby87} & &	 77 & $C_{2v}$ & -409.083517 & \onlinecite{JD95c} \\
 23 & $D_{3h}$ & -92.844472 & \onlinecite{Farges85} & &	 78 & $C_s$ & -414.794401 & $\dagger$ \\
 24 & $C_s$ & -97.348815 & \onlinecite{Wille87} & &	 79 & $C_{2v}$ & -421.810897 & \onlinecite{Northby87} \\
 25 & $C_s$ & -102.372663 & \onlinecite{HoareP71} & &	 80 & $C_s$ & -428.083564 & \onlinecite{Northby87} \\
 26 & $T_d$ & -108.315616 & \onlinecite{HoareP71} & &	 81 & $C_{2v}$ & -434.343643 & \onlinecite{Northby87} \\
 27 & $C_{2v}$ & -112.873584 & \onlinecite{Northby87} & &	 82 & $C_1$ & -440.550425 & \onlinecite{Northby87} \\ 
 28 & $C_s$ & -117.822402 & \onlinecite{Northby87} & &	 83 & $C_{2v}$ & -446.924094 & \onlinecite{Northby87} \\ 
 29 & $D_{3h}$ & -123.587371 & \onlinecite{HoareP71} & &	 84 & $C_1$ & -452.657214 & \onlinecite{Northby87} \\
 30 & $C_{2v}$ & -128.286571 & \onlinecite{Northby87} & &	 85 & $C_{3v}$ & -459.055799 & \onlinecite{Northby87} \\
 31 & $C_s$ & -133.586422 & \onlinecite{Northby87} & &	 86 & $C_1$ & -465.384493 & \onlinecite{Northby87} \\
 32 & $C_{2v}$ & -139.635524 & \onlinecite{Northby87} & &	 87 & $C_s$ & -472.098165 & \onlinecite{Northby87} \\
 33 & $C_s$ & -144.842719 & \onlinecite{Northby87} & &	 88 & $C_s$ & -479.032630 & \onlinecite{Deaven96} \\
 34 & $C_{2v}$ & -150.044528 & \onlinecite{Northby87} & &	 89 & $C_{3v}$ & -486.053911 & \onlinecite{Northby87} \\
 35 & $C_1$ & -155.756643 & \onlinecite{Northby87} & &	 90 & $C_s$ & -492.433908 & \onlinecite{Northby87} \\
 36 & $C_s$ & -161.825363 & \onlinecite{Northby87} & &	 91 & $C_s$ & -498.811060 & \onlinecite{Northby87} \\
 37 & $C_1$ & -167.033672 & \onlinecite{Northby87} & &	 92 & $C_{3v}$ & -505.185309 & \onlinecite{Northby87} \\
 38 & $O_h$ & -173.928427 & \onlinecite{Pillardy,JD95c} & &	 93 & $C_1$ & -510.877688 & \onlinecite{Northby87} \\
 39 & $C_{5v}$ & -180.033185 & \onlinecite{Northby87} & &	 94 & $C_1$ & -517.264131 & \onlinecite{Northby87} \\
 40 & $C_s$ & -185.249839 & \onlinecite{Northby87} & &	 95 & $C_1$ & -523.640211 & \onlinecite{Northby87} \\
 41 & $C_s$ & -190.536277 & \onlinecite{Northby87} & &	 96 & $C_1$ & -529.879146 & \onlinecite{Northby87} \\
 42 & $C_s$ & -196.277534 & \onlinecite{Northby87} & &	 97 & $C_1$ & -536.681383 & \onlinecite{Northby87} \\
 43 & $C_s$ & -202.364664 & \onlinecite{Northby87} & &	 98 & $C_s$ & -543.642957 & \onlinecite{Deaven96} \\
 44 & $C_1$ & -207.688728 & \onlinecite{Northby87} & &	 99 & $C_{2v}$ & -550.666526 & \onlinecite{Northby87} \\
 45 & $C_1$ & -213.784862 & \onlinecite{Northby87} & &	 100 & $C_s$ & -557.039820 & \onlinecite{Northby87} \\
 46 & $C_{2v}$ & -220.680330 & \onlinecite{Northby87} & &	 101 & $C_{2v}$ & -563.411308 & \onlinecite{Northby87} \\
 47 & $C_1$ & -226.012256 & \onlinecite{Northby87} & &	 102 & $C_{2v}$ & -569.363652 & \onlinecite{JD95d} \\
 48 & $C_s$ & -232.199529 & \onlinecite{Northby87} & &	 103 & $C_s$ & -575.766131 & \onlinecite{JD95d} \\
 49 & $C_{3v}$ & -239.091864 & \onlinecite{Northby87} & &	 104 & $C_{2v}$ & -582.086642 & \onlinecite{JD95d} \\
 50 & $C_s$ & -244.549926 & \onlinecite{Northby87} & &	 105 & $C_1$ & -588.266501 & \onlinecite{Northby87} \\
 51 & $C_{2v}$ & -251.253964 & \onlinecite{Northby87} & &	 106 & $C_1$ & -595.061072 & \onlinecite{Northby87} \\
 52 & $C_{3v}$ & -258.229991 & \onlinecite{Northby87} & &	 107 & $C_s$ & -602.007110 & $\dagger$ \\
 53 & $C_{2v}$ & -265.203016 & \onlinecite{Northby87} & &	 108 & $C_s$ & -609.033011 & \onlinecite{Northby87} \\
 54 & $C_{5v}$ & -272.208631 & \onlinecite{Northby87} & &	 109 & $C_1$ & -615.411166 & \onlinecite{Northby87} \\
 55 & $I_h$ & -279.248470 & \onlinecite{HoareP72} & &	 110 & $C_s$ & -621.788224 & \onlinecite{Northby87} \\
 56 & $C_{3v}$ & -283.643105 & \onlinecite{Northby87} & &	
\end{tabular}
\end{table}
\end{minipage}
\begin{multicols}{2}

\begin{minipage}{8.cm}
\begin{table}
\caption{\label{table:nonicos}Lowest energy icosahedral minima at sizes
with non-icosahedral global minima.}
\begin{tabular}{cccc}
 N & Point group & Energy/$\epsilon$ & Ref. \\
 \hline
 38 & $C_{5v}$ & -173.252378 & \onlinecite{Deaven96} \\
 75 & $C_1$ & -396.282249 & \onlinecite{JD95c} \\
 76 & $C_1$ & -402.384580 & \onlinecite{Xue} \\
 77 & $C_1$ & -408.518265 & \onlinecite{Xue} \\
 102 & $C_s$ & -569.277721 & \onlinecite{Northby87} \\
 103 & $C_1$ & -575.658879 & \onlinecite{Northby87} \\
 104 & $C_s$ & -582.038429 & \onlinecite{Northby87} \\
\end{tabular}
\end{table}
\end{minipage}

\acknowledgements
We are grateful to the Royal Society (D.J.W.) and the Engineering and Physical Sciences Research Council 
(J.P.K.D.) for financial support.

\end{multicols}

\begin{thebibliography}{10}
\bibitem{SA}
The magnitude of this interest can be gauged by, for example, the fact that
  there have been over 2600 citations of the classic simulated annealing paper
  by Kirkpatrick \etal, cited below.
\bibitem{pinch}
The exact value of this figure should not be be taken too seriously since it
  depends sensitively on the functional form used to fit Tsai and Jordan's
  results for the number of minima. The value given was obtained using the form
  $\exp(a+bN)$, but if one instead uses the form $\exp(a+bN+cN^2)$ the estimate
  is of the order of $10^{259}$ minima.
\bibitem{dcoex}
For the larger sizes coexistence in the melting region may occur between the
  liquid-like state and defective structures based on the global
  minimum,\cite{Kunz93,Kunz94} and hence molecular dynamics runs at two
  energies would probably be necessary, first in the melting region and then at
  a lower energy starting from one of the defective icosahedral structures
  generated.
\end{thebibliography}
\end{document}